\documentclass[aps,prb,twocolumn,nopacs,superscriptaddress]{revtex4-1}

\usepackage{amsmath,amssymb,graphicx}

\pdfoutput=1
\usepackage{hyperref}
\hypersetup{
   pdfauthor={KSD Beach and FF Assaad},
   pdftitle={Orbital-selective Mott transition and heavy fermion behavior in a bilayer Hubbard model for 3He.}
}

\newcommand{\bracket}[1]{\langle #1 \rangle}
\renewcommand{\vec}[1]{\mathbf{#1}}

\DeclareMathOperator{\Imag}{Im}
\DeclareMathOperator{\Tr}{Tr}

\begin{document}

\title{Orbital-selective Mott transition and heavy fermion behavior\\ 
in a bilayer Hubbard model for $^3$He}

\author{K. S. D. Beach}
\affiliation{Department of Physics, University of Alberta, Edmonton, Alberta, Canada T6G 2G7}

\author{F. F. Assaad}
\affiliation{Institut f\"ur theoretische Physik und Astrophysik,
Universit\"at W\"urzburg, Am Hubland, D-97074 W\"urzburg, Germany}

\begin{abstract}
Inspired by recent experiments on $^3$He films between one and two atoms thick, 
we consider a bilayer Hubbard model on a
triangular lattice. Our results are obtained in the framework of a cluster dynamical 
mean-field calculation with a quantum Monte Carlo impurity solver.
For appropriate model parameters, we observe an enhancement of the effective mass as 
the first layer approaches integer filling and the second remains only partially filled. At finite 
temperatures, this increase of the effective mass---or, equivalently, the decrease of the coherence 
temperature---leads to a crossover to a state where the first layer fermions localize, drop out of the Luttinger volume, 
and generate essentially free local moments. This finite temperature behavior is shown to be robust against 
the cluster size above some critical temperature. The zero temperature phase diagram, however, depends 
on the cluster topology. In particular, for clusters with an even number of unit cells, the growth of the effective mass is cut off
by a first-order, orbital-selective Mott transition.
\end{abstract}


\maketitle

\section{Introduction} 

The solidification of $^3$He monolayers~\cite{Casey03} has been 
interpreted as a density-driven Mott transition in which the effective 
mass diverges.~\cite{Vollhardt_rev,Imada_rev} Below the critical density, 
the system is a metallic, nearly localized Fermi liquid; beyond the critical density, 
it is a solid, the magnetic properties of which are dominated by antiferromagnetic 
two-body exchange processes.~\cite{Thouless65,Roger83,Elser89}
It is now possible to realize bilayers of $^3$He (atop a frozen $^4$He substrate, itself adsorbed onto graphite) 
with the special property that the second layer begins to form before the first has solidified.~\cite{Neumann07}
Since the first layer is close to a Mott transition, the $^3$He fermions in 
this layer are slow (i.e.\ heavy), whereas those in the second layer are fast. 
This combination of fast and slow dynamics---corresponding to wide and narrow fermion conduction
bands---is completely analogous to the situation in electronic heavy fermion materials, albeit
without the complication of crystal field and spin orbit effects.

According to this picture, one expects, prior to solidification of the first layer, an 
enhanced effective mass and a Luttinger volume that counts both the first- and 
second-layer populations. Moreover, one naively anticipates that further $^3$He
deposition will eventually cause the effective mass to diverge, in coincidence with the solidification of the 
first layer. This solidification of the first layer can be interpreted either as an orbital-selective Mott 
transition or, in the terminology of Kondo physics, as a Kondo breakdown in which the heavy particles 
drop out of the Luttinger volume. 
In experiment, the effective mass is indeed observed to increase as a function of the 
total $^3$He concentration, but its growth is interrupted by an intervening phase.~\cite{Neumann07}
The fact that this phase is ferromagnetic indicates that three-body 
exchange processes come to dominate in the solid phase of the first layer.~\cite{Thouless65,Roger83}

\begin{figure}
\includegraphics{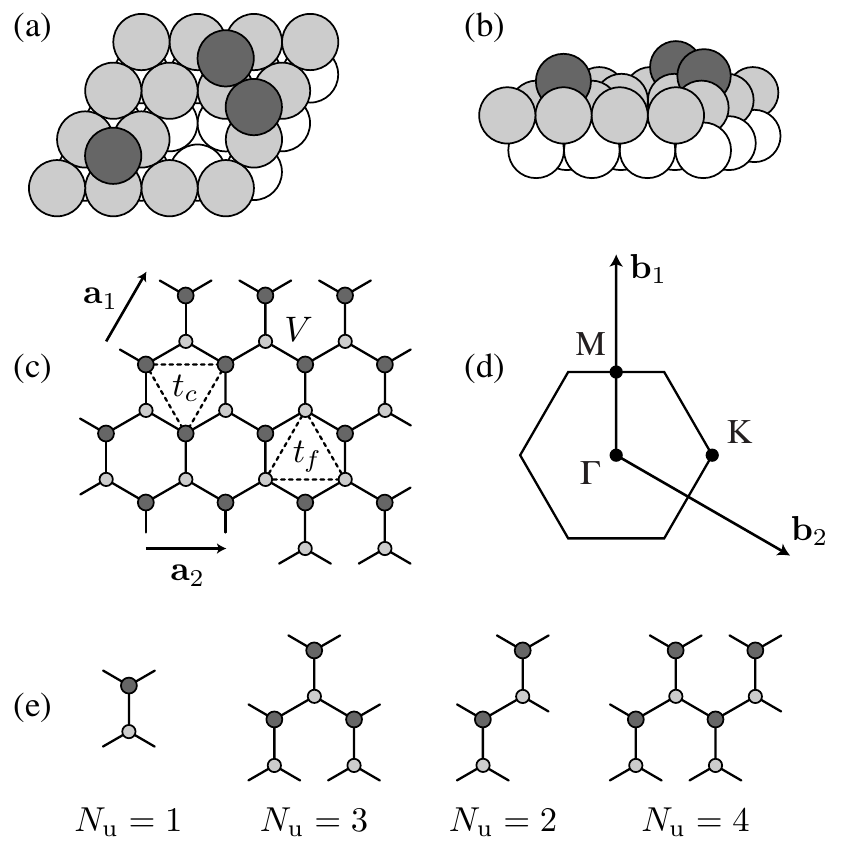} 
\caption{
(a),(b)~Stacking of billiard balls modeling of bilayer $^3$He, top and side view, with 
the $^4$He substrate shown in white.
(c)~Tight-binding modeling with hoppings $t_c$, $t_f$, and $V$.
(d)~The hexagonal Brillouin zone of the triangular lattice.
(e)~The set of supercells, each consisting of $N_{\text{u}}$ unit cells, considered in this work. 
\label{model.fig} }
\end{figure}
 
The motivation of this article is to consider a simple lattice model that goes a 
good way towards reproducing the essential features of the above experimental situation. 
As shown in Figs.~\ref{model.fig}(a) and \ref{model.fig}(b), we adopt a {\it stacking of billiard balls} 
modeling of 
bilayer $^3$He on a triangular lattice defined 
by $\vec{a}_1 =(1/2, \sqrt{3}/2,0)$ and $\vec{a}_2 = (1,0,0)$. Each unit cell 
accounts for two $^3$He positions, $\vec{r}_f = \vec{0}$ and $\vec{r}_c = 
\frac{2}{3} \vec{a}_1 - \frac{1}{3} \vec{a}_2 + (0,0,a_3)$, measured relative to the lattice.
This geometry presupposes a particular stacking arrangement for the second $^3$He layer.

Our model can be viewed as a honeycomb lattice whose inequivalent sites (corresponding
to $^3$He positions in the upper and lower layers) are populated by two species of fermion, 
which we label $c$ and $f$.
The tight-binding parameters include a nearest-neighbor (interlayer) hopping $V$ and next-nearest-neighbor
(intralayer) hoppings $t_c$ and $t_f$. See Fig.~\ref{model.fig}(c). With the inclusion of onsite Coulomb repulsion terms, 
the Hamiltonian reads
\begin{equation} 
\label{Model.Eq}
\begin{gathered}
 H = \sum_{\vec{k},\sigma} 
 \begin{pmatrix} c^{\dagger}_{\vec{k},\sigma} & f^{\dagger}_{\vec{k},\sigma} \end{pmatrix}
 \begin{pmatrix}
 \varepsilon_c(\vec{k}) - \mu & V(\vec{k}) \\
 V(\vec{k}) & \varepsilon_f(\vec{k}) - \mu
 \end{pmatrix}
 \begin{pmatrix}
 c_{\vec{k},\sigma} \\ f_{\vec{k},\sigma}
 \end{pmatrix} \\
{} + U_c \sum_{\vec{i}} \bigl( \hat{n}_{c,\vec{i}} - 1 \bigr)^2 
 + U_f \sum_{\vec{i}}\bigl( \hat{n}_{f,\vec{i}} - 1 \bigr)^2. 
\end{gathered}
\end{equation} 
Here, the mixing element $V(\vec{k}) = V(3+2\gamma_{\vec{k}})^{1/2}$ and the
dispersion $\varepsilon_c(\vec{k}) = -2 t_c \gamma_{\vec{k}} + \varepsilon_c^{0}$
are expressed
in terms of the connection
$\gamma_{\vec{k}} = \cos(\vec{k}\cdot \vec{a}_1) + \cos(\vec{k}\cdot \vec{a}_2) + 
\cos[\vec{k}\cdot (\vec{a}_2-\vec{a}_1)]$
of the underlying Bravais lattice. The operator
$\hat{n}_{c,\vec{i}} = \sum_{\sigma} c^{\dagger}_{\vec{i},\sigma}
c_{\vec{i},\sigma}$ is the local $^3$He density in the upper layer.
Similar definitions hold for $\varepsilon_f(\vec{k})$ and $\hat{n}_{f,\vec{i}}$.

Except for the complication of the layer stacking (and the resulting $\vec{k}$-dependent
hybridization), this bilayer Hubbard model reduces to the Periodic Anderson Model 
as $t_f \rightarrow 0$, a limit in which the bare mass of the $f$ fermions diverges. 
Similar models have been considered for the description of bilayer $^3$He in Refs.~\onlinecite{Benlagra08}
and \onlinecite{Benlagra09}
within a slave boson mean-field calculation. Here we go a significant step further and 
perform calculations within the cellular dynamical mean field theory (CDMFT)~\cite{Biroli04} 
approximation (Sec.~\ref{CDMFT}). 
Our strategy is to systematically investigate the model of Eq.~(\ref{Model.Eq}) as a function of the cluster size. 
In Sec.~\ref{Results}, we will see that clusters with an odd number of unit cells have a radically different 
low-energy behavior than those with an even number. Given this situation, the extrapolation to the 
large cluster size limit is delicate and is relegated to the conclusions in Sec.~\ref{Conclusions}. 
Part of this work has already appeared in a preprint.~\cite{Beach09}

\section{Cellular Dynamical Mean Field Theory\label{CDMFT}}

By construction, the CDMFT approach exactly accounts for the temporal fluctuations at each site 
and thereby captures the physics of the local moments---both their formation 
and their screening via the Kondo effect. But the spatial fluctuations extend only over the simulation cluster;
insofar as the true correlation length scale exceeds the linear size of the cluster, the results will suffer from finite size effects. 
To mitigate this, we have considered various cluster topologies ranging from one unit cell 
(a single $c$ and $f$ site) to four unit cells as 
defined in Fig.~\ref{model.fig}(e). For a given supercell, the resulting single particle Green function,
 $\underline{G} ( \vec{K}, i \omega_m ) $, is a $2 N_{\text{u}} \times 2 N_{\text{u}} $ matrix 
with crystal momentum $\vec{K}$ in the Brillouin zone of the supercell lattice.
The CDMFT calculation involves neglecting momentum conservation and 
thereby obtaining a $\vec{K}$-independent self-energy 
$\underline{\Sigma}(i \omega_m) $. This quantity is extracted from a cluster of 
$N_{\text{u}}$ unit cells embedded in a dynamical mean field that is determined self-consistently. We have
solved this cluster problem using a standard Hirsch-Fye approach and have symmetrized the cluster 
Green function to obtain the corresponding quantity on the lattice:
\begin{equation}
\label{Greenk.eq}
 G(\vec{k},i\omega_m)_{\mu,\nu} = \frac{1}{N_{\text{u}}} \sum_{\alpha,\beta} e^{i \vec{k} \cdot
 \left( \vec{x}_{\alpha} - \vec{x}_{\beta} \right)} 
 \underline{G} ( \vec{K}, i \omega_m )_{(\mu,\alpha), (\nu, \beta)}. 
\end{equation}
Here $\vec{x}_{\alpha}$ denotes the unit cell positions within the supercell, 
$\mu$ and $\nu$ run over the $c$ and $f$ orbitals within each unit cell, 
and $\vec{k}$ and $\vec{K}$ differ by a reciprocal lattice vector of the supercell 
Bravais lattice. The rotation to real frequencies was carried out with 
a stochastic analytical continuation technique.~\cite{Sandvik98,Beach04a}

\section{Results\label{Results}}

We consider the following model parameters: $t_c = t_f = t$, $U_c/t = U_f/t = 12$, 
$V/t = 1/2$, $\varepsilon^0_c/t = 3$, and $\varepsilon^0_f/t = 0$. We have chosen large values 
of $U_c$ and $U_f$ to reflect the contact repulsion of the $^3$He atoms and to guarantee 
that each single layer is well within the Mott insulating phase at half-band filling.~\cite{Kyung07}
These values of the Hubbard interaction lead to low double occupancy,
thus generating local moments. The difference $\varepsilon^0_c - \varepsilon^0_f > 0$ is a
crude accounting for the van der Waals forces (both $^4$He--$^3$He and $^3$He--$^3$He)
that preferentially fill the first layer. 

\begin{figure*}
 \includegraphics{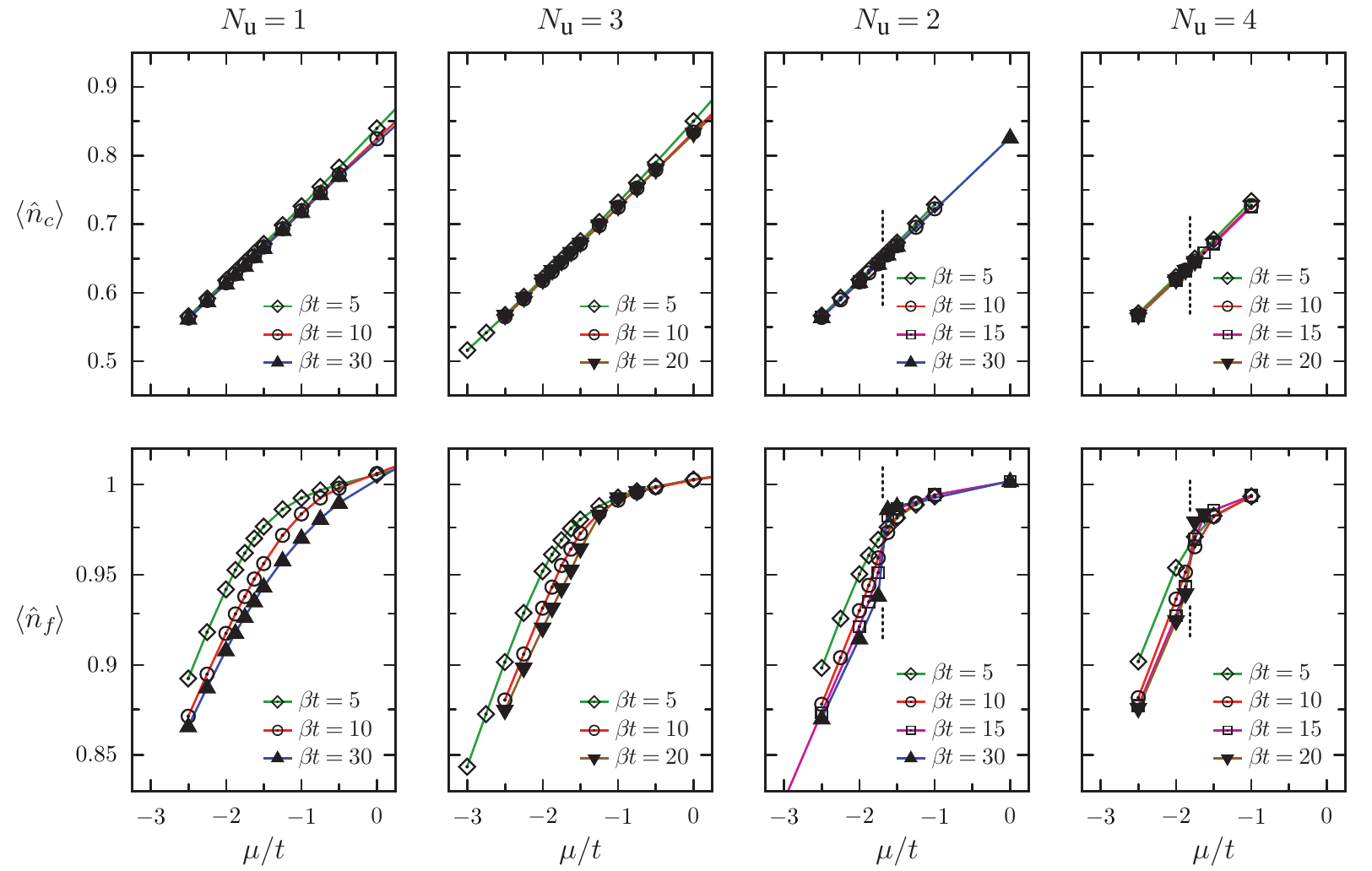} 
\caption{
\label{chem.fig}
The average occupation number is plotted for the upper- (top row) and lower-layer (bottom row) fermions 
as a function of chemical potential. Results are reported for both odd- (leftmost two columns)
and even-numbered (rightmost two columns) clusters. Solid lines connect data sets with a common
temperature. Vertical dashed lines mark the location of the emerging low-temperature discontinuity
in $\bracket{\hat{n}_f}$.
}
\end{figure*}

\subsection{Layer densities}
The generic Mott insulating state is characterized by a density $\bracket{\hat{n}} = 1$ and a vanishing charge 
susceptibility: i.e., $ \chi_{\text{ch}} = \partial\bracket{\hat{n}}/\partial \mu = 0$, where $\mu$ denotes the chemical 
potential. Figure~\ref{chem.fig} plots the layer-resolved 
densities $\bracket{\hat{n}_c}$ and $\bracket{\hat{n}_f}$ as a function of the chemical potential,
which controls the overall $^3$He concentration. For both the odd and even cluster sizes,
$\bracket{\hat{n}_f}$ shows a plateau feature centered around $\bracket{\hat{n}_f} = 1$, whereas 
$\bracket{\hat{n}_c} $ grows smoothly. In contrast to the Mott insulating state, 
$ \chi_{\text{ch}}^f = \partial \bracket{\hat{n}_f}/\partial \mu$ never vanishes. Hence, charge fluctuations between 
the layers are allowed and the simple picture of a complete 
decoupling of the layers never holds. Although the plateau feature is common to all cluster sizes, the data show
distinct odd-even effects. 
For $N_{\text{u}} = 1$ and $N_{\text{u}}=3$, $\bracket{\hat{n}_f}$ is a continuous function of the chemical potential for all 
temperatures considered. In contrast, for the even clusters, $N_{\text{u}} = 2$ and $N_{\text{u}} = 4$, a discontinuity in 
$\bracket{\hat{n}_f}$ emerges below a critical temperature $T_c$ and at a critical chemical potential. For
$N_{\text{u}} = 4$, a robust discontinuity is present at $T_c \simeq t/20$, whereas for $N_{\text{u}}=2$ this feature 
already appears at $ T_c \simeq t/15$. 
Since $\bracket{\hat{n}_c + \hat{n}_f} = \partial F/\partial \mu$, where $F$ is the free energy, the jump 
in the total fermionic density signals a density-driven first-order transition. 
In a canonical ensemble, states with total density lying within the jump are phase separated. 

\subsection{First-layer effective mass and low temperature spectral functions}

\begin{figure}
\includegraphics{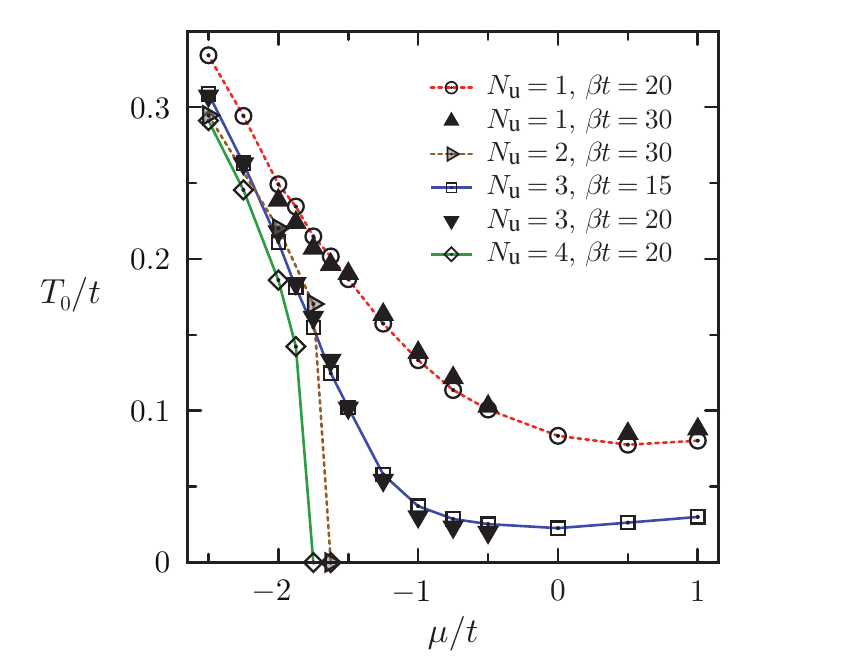} 
\caption{\label{residue.fig} The inverse of quantity defined in Eq.~\eqref{Z.eq}, which is proportional to
the coherence temperature $T_0$, inverse effective mass, and 
quasiparticle residue, is plotted as a function of chemical potential. }
\end{figure} 

We can estimate the $f$ fermion's effective mass as a function of the chemical potential by considering its 
cluster-averaged self-energy, 
$\Sigma_{f}(i\omega_m) = \frac{1}{N_{\text{u}}} \sum_{\alpha= 1}^{N_{\text{u}}} \Sigma_{(f,\alpha),(f,\alpha)} (i \omega_m)$,
and extracting the quantity
\begin{equation}
\label{Z.eq}
 \frac{t}{T_{0}}=\frac{m^{\star}}{m} \propto Z^{-1} = 1 - \frac{ \Imag \Sigma_{f}(i \omega_m = i \pi T)}{\pi T }.
\end{equation}
This estimate of the effective mass (or, equally, of the inverse of the coherence temperature $T_0$) 
is valid provided that the real space dependence of the self-energy is small and
that the temperature $T$ is extrapolated to zero. Data on the $N_{\text{u}} = 4$ cluster presented in 
Ref.~\onlinecite{Beach09} shows that for $\mu < \mu_c $ the assumption of a local self-energy is valid. 

\begin{figure}
 \includegraphics{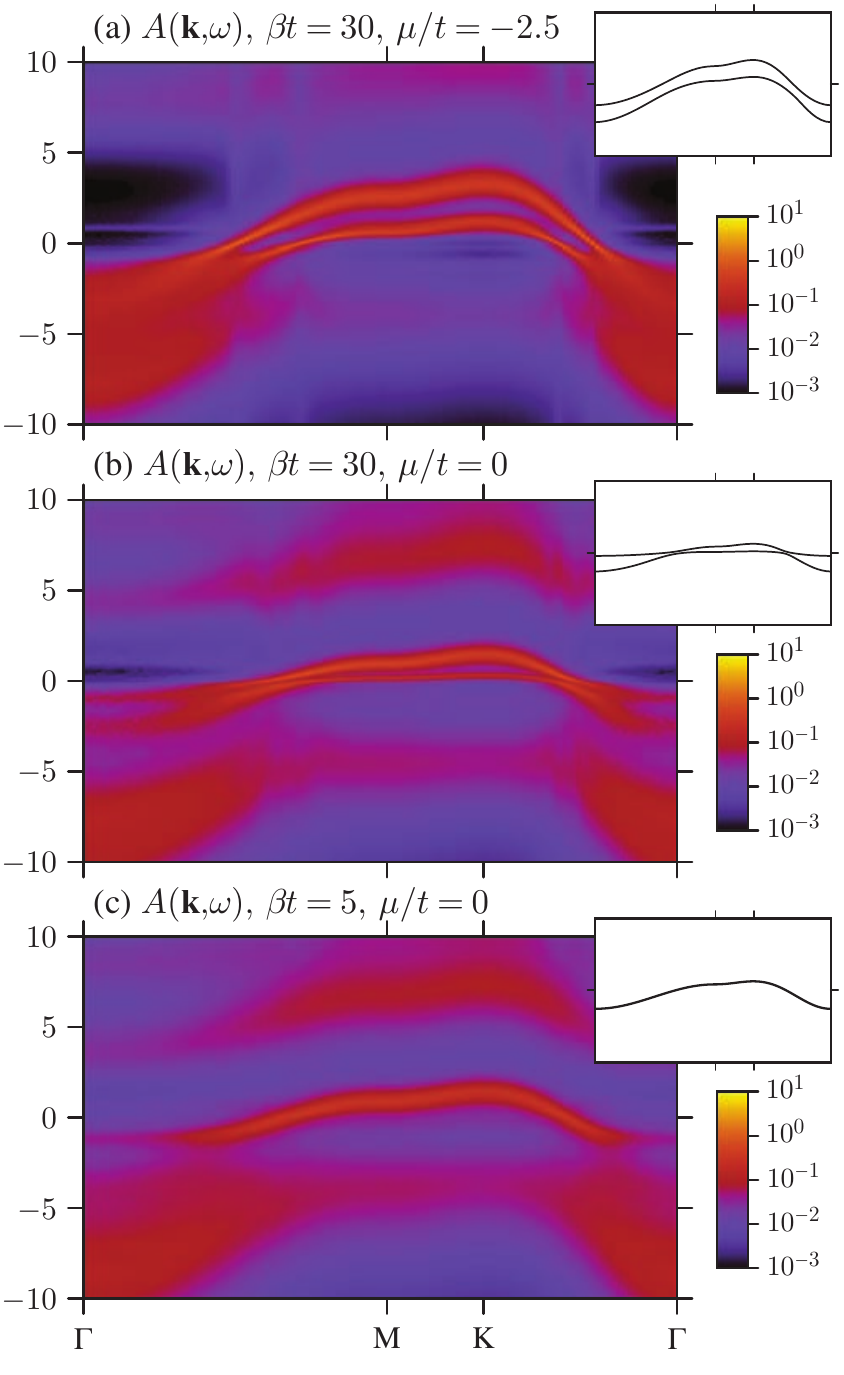} 
\caption{ \label{Low_T_Akom_Nu_1.fig} 
The amplitude of the single particle spectral function,
as defined by Eq.~\eqref{Akom.eq}, is plotted for the $N_{\text{u}}=1$ cluster
at various values of the chemical potential 
and temperature. The insets show the corresponding to slave-boson calculation.
The inset to panel (a) shows the mean-field band structure consisting
of two strongly-hybridized quasiparticle bands of mixed $c$ and $f$ character. 
In panel (b), the hybdridization is considerably weakened. In panel (c), the $f$
fermions have dropped out of the Luttinger volume.
}
\end{figure} 

At $N_{\text{u}}=1$ (see Fig.~\ref{residue.fig}), the effective mass increases as a function of chemical potential.
This effect is also evident in the evolution of single particle spectral functions, 
\begin{equation}
\label{Akom.eq}
A(\vec{k},\omega) = - \Imag \Tr G(\vec{k},\omega+i0^+), 
\end{equation} 
plotted in Fig.~\ref{Low_T_Akom_Nu_1.fig}.
As exemplified by the data set at $\mu/t = -2.5$ [Fig.~\ref{Low_T_Akom_Nu_1.fig}(a)], the low-energy 
coherent features of the spectral function compare favorably with a slave 
boson approximation leading to mass-renormalized hybridized bands. This state has a Luttinger volume that includes both 
$f$ and $c$ fermions, and the band with the largest Fermi volume has dominant $f$ character. As a function of 
the chemical potential, the effective mass of the $f$ band grows, and spectral weight is shifted to the upper Hubbard band. 
At $\mu =0$ and $\beta t = 30$, the data of Fig.~\ref{Low_T_Akom_Nu_1.fig}(b) exhibits typical heavy fermion character: 
a lower Hubbard band located at $\omega_L /t \simeq -6 $, an upper Hubbard band at $ \omega_U/t \simeq 6 = \omega_L/t + U_f/t$, and 
a heavy band with dominant $f$ character in close vicinity of the Fermi energy that hybridizes with a light conduction band. At $N_{\text{u}}=3$, the coherence temperature (see Fig.~\ref{residue.fig}) is reduced with respect to the 
$N_{\text{u}}=1$ case but nevertheless shows a similar 
overall behavior: a rapid decrease as a function of chemical potential followed by saturation at a lower value than for the 
$N_{\text{u}}=1$ case. Within the accessible temperature range of the $N_{\text{u}}=3$ cluster, 
the single particle spectral function shows the same features as for the $N_{\text{u}}=1$ case. 
 
\begin{figure}[b]
\includegraphics{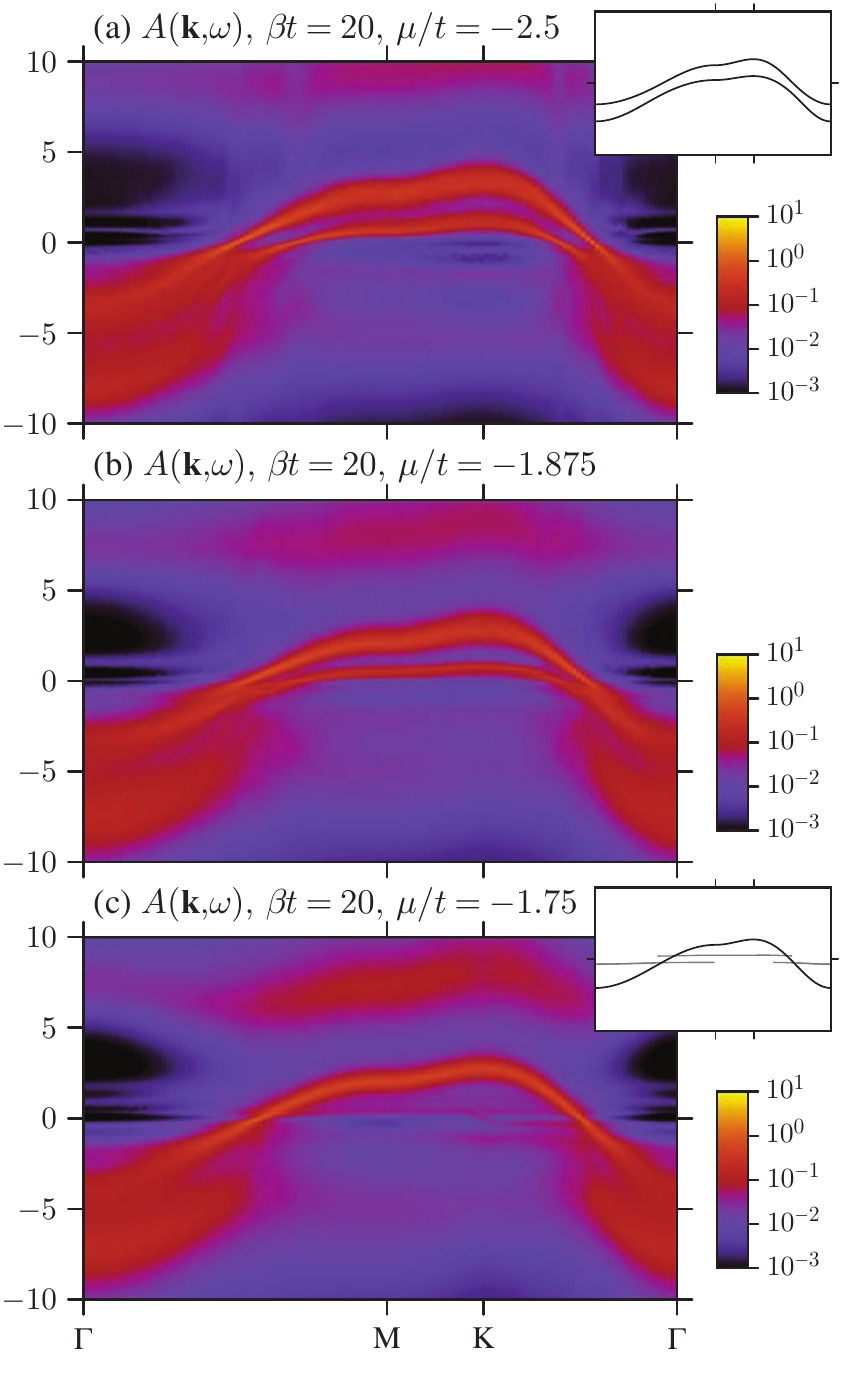} 
\caption{\label{Akom.fig} The amplitude of the single particle spectral function
for the $N_{\text{u}}=4$ cluster. The inset in panel (c) shows the single $c$-only band completely
decoupled from the gapped, nearly flat band of the singlet-bound $f$ fermions. The mean-field calculations were carried 
out on the $N_{\text{u}}=2$ cluster and the QMC on the $N_{\text{u}}=4 $ systems. 
}
\end{figure}

For the even site clusters, the initial decrease of the coherence temperature is cut off by the first-order transition. 
Far below $\mu_c$, as exemplified by $\mu/t = -2.5$, the single particle spectral function is 
very similar to that observed on the $N_{\text{u}}=1$ cluster [cf.\ Figs.~\ref{Low_T_Akom_Nu_1.fig}(a)
and \ref{Akom.fig}(a)]. 
With increasing chemical potential, the effective mass of the $f$ band grows, and beyond $\mu_c$ the $f$ band drops 
out of the low-energy physics altogether. This can be understood at the \emph{static}
mean field level by a conventional slave boson theory in competion with local singlet 
formation in the first layer. The transition is signaled by the appearance of an 
anomalous expectation value $\Delta_{\vec{i}\vec{j}} \sim (t_f^2/U_f)\sum_{\sigma} \bracket{f_{\vec{i},\sigma}^\dagger f_{\vec{j},\sigma}}$. The inset of Fig.~\ref{Akom.fig}(c) shows the band structure that results 
when this singlet order parameter breaks down the original lattice symmetry to that of the $N_{\text{u}}=2$ supercell. 

\subsection{Spin susceptibilities and correlations }

\begin{figure}
\includegraphics{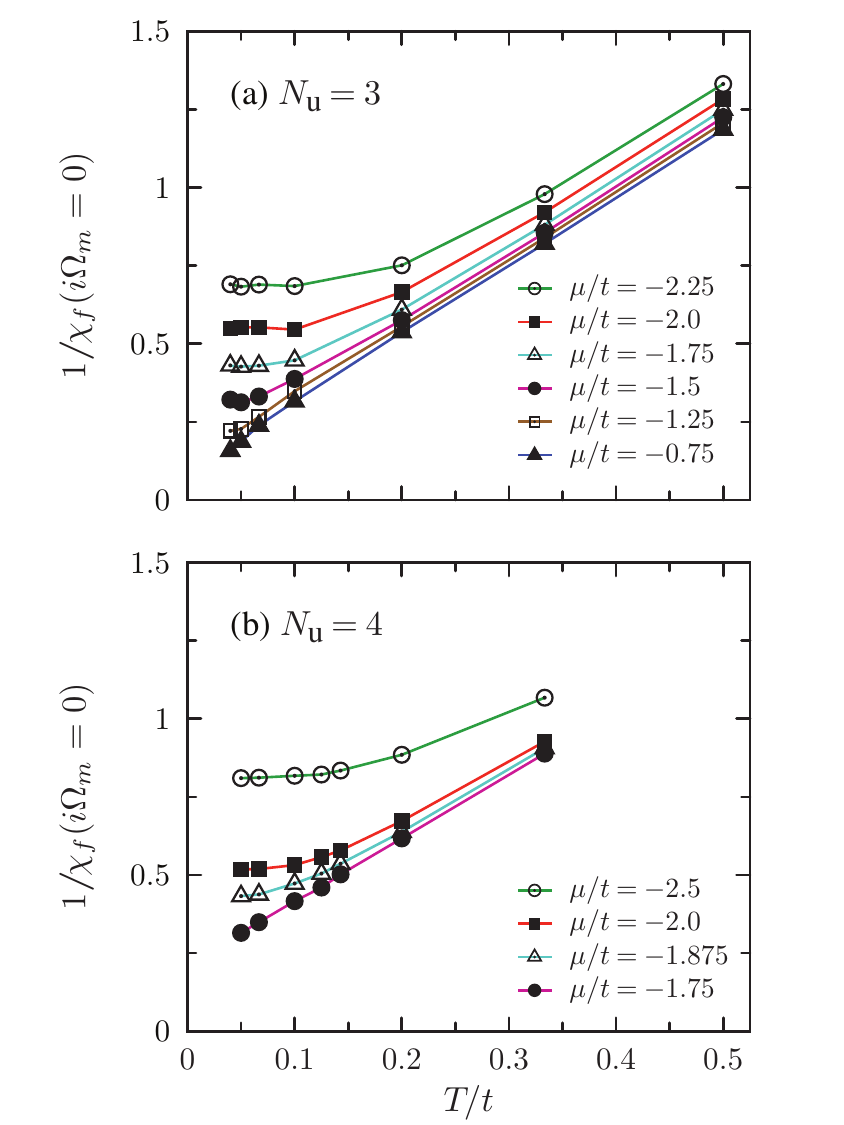} 
\caption{\label{Spin_sucep.fig}
Temperature dependence of the inverse local spin susceptibility for the (a)~$N_{\text{u}}=3$ and 
(b)~$N_{\text{u}}=4$ clusters.} 
\end{figure}

We can extract from the cluster the local spin susceptibility as defined by 
\begin{equation}
 \chi_f(i \Omega_m) = \frac{1}{N_{\text{u}}} \sum_{\vec{i}} \int_{0}^{\beta} \!{\rm d} \tau\,
 e^{i \Omega_m \tau} \langle \vec{S}^f_{\vec{i}}(\tau) \cdot \vec{S}^f_{\vec{i}}(0) \rangle.
\end{equation}
A Fermi liquid below its Fermi temperature is Pauli paramagnetic, and hence $\chi_f(i \Omega_m=0)$ is constant.
On the other hand, a local moment is characterized by a
Curie-Weiss law, $ \chi_f(i \Omega_m=0) \propto 1/(T + \Theta)$ at temperatures $T \gg \Theta$.
As is apparent in Fig.~\ref{Spin_sucep.fig}, $\chi_f(i \Omega_m=0)$ always exhibits a smooth crossover from 
the high-temperature Curie-Weiss to the low-temperature Pauli behavior, irrespective of the lattice topology. 
For the odd lattice sizes, the crossover point tracks the 
coherence temperature. The same holds for the even lattice at $\mu < \mu_c$. It is worth emphasizing that
this qualitative change in magnetic response pinned to the coherence temperature (also denoted by $T_0$ 
in Ref~\onlinecite{Neumann07}) 
has been observed in the $^3$He bilayer experiment.~\cite{Neumann07}
Hence, at high temperatures, a local moment generated by the Hubbard interaction is present. 
The screening of this local moment, or in other words the quenching of its entropy, is at the origin of 
the different behavior between the odd- and even-numbered lattices. \\
\underline{$N_{\text{u}}=1$.} For this cluster smallest size, only the delocalized $c$ fermions are available to
screening the local moment. This is precisely the Kondo effect, and one can view the heavy fermion paramagnetic state 
as originating from the coherent, Bloch-like superposition of individual Kondo screening clouds. Within a periodic 
Anderson model, this screening of the local moment is linked to a delocalization of the $f$ fermion. Hence, 
above $T_{0}$, when screening is absent, we expect the 
$f$-quasiparticle band to drop out of the low-energy physics. 
This is evident from Figs.~\ref{Low_T_Akom_Nu_1.fig}(b) and \ref{Low_T_Akom_Nu_1.fig}(c) 
upon comparison of the high- and low-temperature spectral 
functions at $\mu = 0$. 
\begin{figure}
\includegraphics{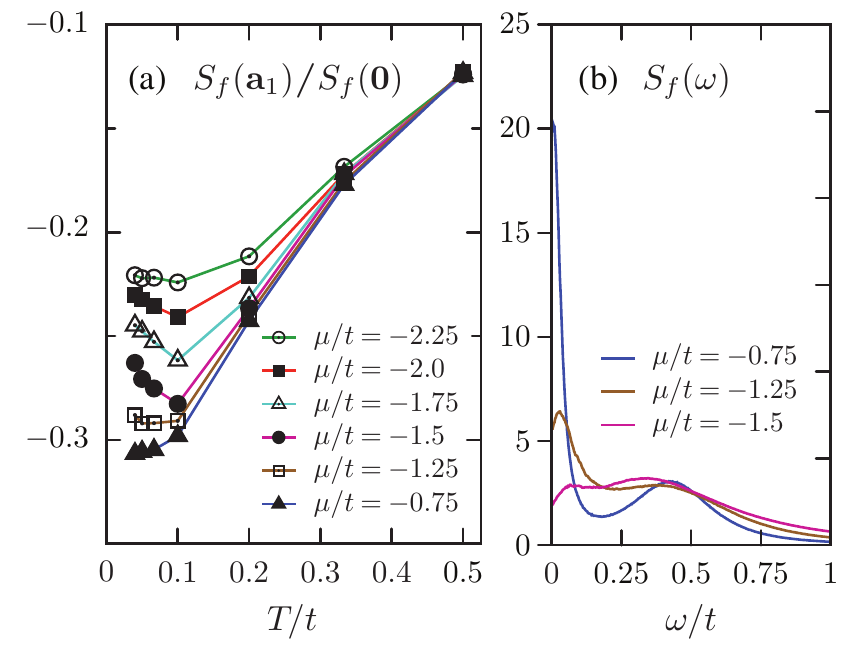} 
\caption{\label{Spin-Spin_Nu3.fig}
(a)~Normalized nearest-neighbor $f$-fermion spin-spin correlations for on the $N_{\text{u}}=3$ cluster. On a three-site ring, the value of this quantity for the Heisenberg model is given by $-1/3$. (b)~Local dynamical spin structure factor. } 
\end{figure}\\
\underline{$N_{\text{u}}=3$.} This cluster size shows behavior very similar to that of the $N_{\text{u}} = 1$ system, albeit with a lower 
coherence temperature. At values of the chemical potential where the $f$ layer is approximately half filled, 
 a magnetic superexchange interaction $ J = 4 t^2 /U_f $ is dynamically generated and the spin degrees of freedom 
on the first layer are described by a 
Heisenberg model on a three-site lattice. The ground state is fourfold degenerate corresponding to a spin-1/2 degree of freedom
with either positive or negative chirality. As in the $N_{\text{u}}=1$ case, the only way to quench this residual entropy is via 
Kondo screening by the $c$ fermions. 
To confirm this interpretation, we have computed the nearest-neighbor spin-spin correlations 
on the first layer, $ S_f(\vec{r}) = \langle \vec{S}^f_{\vec{i}} \cdot \vec{S}^f_{\vec{i} + \vec{r}} \rangle $. 
Comparison with the Heisenberg result is best achieved by normalizing the QMC data by the magnitude of the local moment, 
$ S^f(\vec{r}=\vec{0}) $. As shown in Fig.~\ref{Spin-Spin_Nu3.fig}(a),
the nearest-neighbor antiferromagnetic spin-spin correlations are considerable. 
At high temperatures the energy scale at which they 
decay is set by the superexchange coupling $J/t = 1/3$; and at $\mu/t = -0.75$, where we observe a Curie-Weiss law 
down to our lowest temperature, they compare favorably to the Heisenberg ground state result: 
$ S^f(\vec{a}_1)/ S^f(\vec{0}) = -1/3 $.
Fig.~\ref{Spin-Spin_Nu3.fig}(b) plots the dynamical local spin structure factor,
\begin{equation}
 S_f(\omega) = \Imag \frac{\chi_f(\omega)}{1-e^{-\beta \omega}},
\end{equation}
at $\beta t = 25 $ and as a function of the chemical potential.
As mentioned above, at $\mu /t = -0.75 $ the residual entropy is not quenched. Consequently, a low frequency 
sharp feature in $S_f(\omega)$ marks the spin degenerate ground state of the three-site, half-filled 
Hubbard model. A feature at $\omega/t \simeq 0.5$ corresponds to the first spin excitation, which for the 
three-site Hubbard model at $U_f/t=12 $ takes the value $\Delta_{\text{sp}}/t = 0.49$.
As we decrease the chemical potential from $ \mu/t = -0.75 $ to $\mu/t = -1.5 $, the weight in the high-energy feature 
remains approximately constant, 
but the sharp low-energy feature decreases in intensity and is shifted to slightly higher energies. 
This screening of the residual entropy by the $c$ fermions competes with the nearest-neighbor 
antiferromagnetic fluctuations in the first layer and is at the origin of the upturn in 
$ S^f(\vec{a}_1)/ S^f(\vec{0}) $ (see Fig.~\ref{Spin-Spin_Nu3.fig}) at low temperatures. 

\begin{figure}
\includegraphics{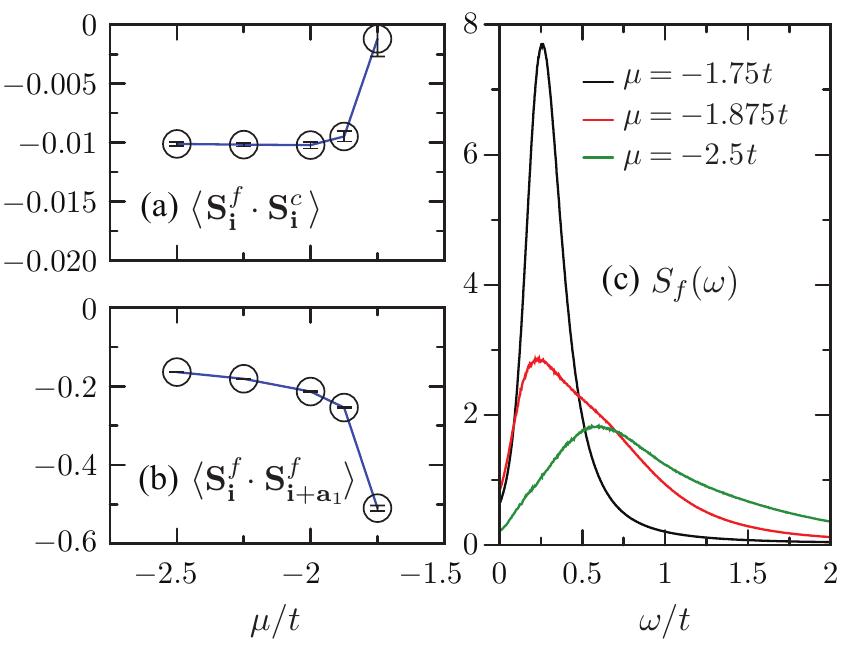} 
\caption{\label{Spin_eq.fig}
 Static and dynamical spin correlations on the $N_{\text{u}}=4$ cluster at $\beta t = 20$.
 (a)~Intracell spin-spin correlations between $c$ and $f$ fermions. 
 (b)~Nearest-neighbor spin-spin correlations between $f$ fermions. 
 (c)~Local dynamical spin structure factor. 
} 
\end{figure}

Hence, at $N_{\text{u}} =3$ spin correlations between the $f$ fermions quench part of the entropy associated with the formation of the local moments. The residual entropy is Kondo screened by the $c$ fermions, and in comparison to the $N_{\text{u}}=1$ case leads to a suppressed coherence temperature.\\
\underline{$ N_{\text{u}}=2, N_{\text{u}} = 4$ } The even site clusters show a band-selective Mott transition and a 
low-energy decoupling of the first and second layers. As in the $N_{\text{u}} = 3 $ case, we can consider the effective 
Heisenberg model on the first layer. For even cluster sizes the ground state is unique and is spin singlet. 
 The first-order transition we observed in Fig.~\ref{chem.fig} arises from
competing screening mechanisms of the local moments generated by the nearly localized $f$ fermions. On the
one hand, the local moments can be Kondo screened by the light $c$ fermions, thereby generating heavy fermion 
behavior. On the other hand, they can form (among themselves) a spin singlet state entirely in the first layer. 
The {\it gapping} of the spin and charge degrees of freedom of the $f$ quasiparticles at $\mu > \mu_c$ allows 
for the decoupling of $f$ and $c$ quasiparticles: a $c$ quasiparticle at the Fermi level cannot scatter 
off an $f$ quasiparticle due to the absence of phase space. 
To support the picture of a sudden change in the screening mechanism, 
we plot in Fig.~\ref{Spin_eq.fig} intra- and interlayer equal-time spin-spin correlations for the $N_{\text{u}}=4$ cluster size. 
At the critical chemical potential, we observe a {\it sudden} growth of the antiferromagnetic correlations 
between nearest-neighbor $f$ fermions and a decrease in the intracell $c$-$f$ spin-spin correlations. 
Figure~\ref{Spin_eq.fig} also shows the local dynamical spin structure factor.
One observes a depletion of spectral weight at low energies on both 
sides of the transition and a considerable sharpening of the line shape in the band-selective 
Mott insulating state. At $\mu < \mu_c$, we can interpret the data within an itinerant fermion picture
where the mass enhancement prior to the band-selective Mott transition is taken into account 
by a renormalization of the hybridization $V$ and hopping $t$ as in a slave boson approach.~\cite{Benlagra08}
Following this modeling, the peak position in $S_f(\omega) $ 
is expected to track the coherence temperature or, equivalently, the inverse effective mass. 
An explicit comparison of those quantities is provided in Ref.~\onlinecite{Beach09}.
At $\mu > \mu_c$, $S_f(\omega)$ should be interpreted within a localized $f$ fermion picture, in which case 
the peak position is a measure of the excitation energy required to break the singlet state of the $f$ fermions. 
On a four-site Hubbard cluster, corresponding to the $f$ layer in the $N_{\text{u}} = 4$ case, this quantity is given 
by $0.214 t $ and compares favorably to the data in Fig.~\ref{Spin_eq.fig}. 
 
\section{Conclusions\label{Conclusions} } 

\begin{figure}
\includegraphics{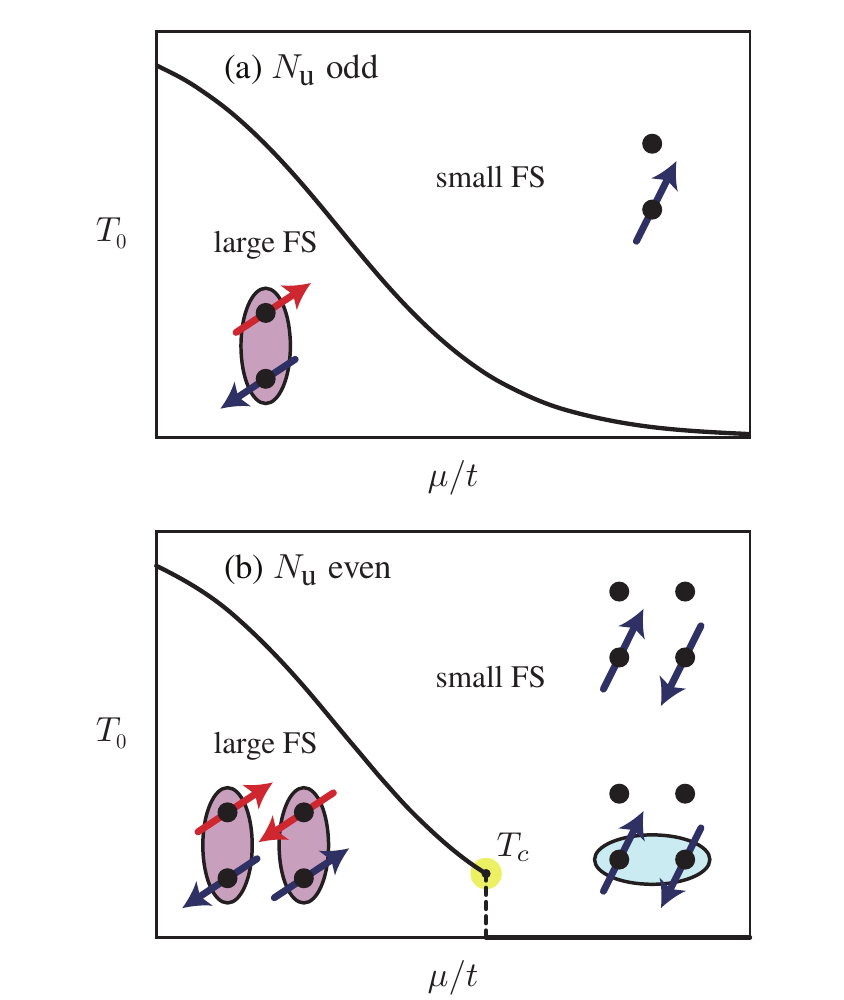} 
\caption{\label{Conc.fig}
Schematic phase diagrams for clusters that are (a) odd and (b) even in number.
The solid line corresponds to the coherence temperature $T_0$, which sets the crossover scale
between states with large and small Fermi surfaces. 
The dashed line is a true phase boundary and corresponds to a line of first-order transitions 
terminating at a critical end point $T_c$. } 
\end{figure}

Our calculations are best summarized by the phase diagrams plotted in Fig.~\ref{Conc.fig}. 
At {\it high} temperatures ($ T > T_c $) the results are independent of the cluster size and topology. 
As the first layer approaches half band-filling the Hubbard $U$ generates a large effective mass as well as 
 local moments. The local moment is 
the key feature of the high temperature phase and results in a Curie-Weiss spin susceptibility. 
In the bilayer $^3$He experiment this is indeed observed at temperatures above $T_0$ prior to the solidification of 
the first layer. This high temperature local moment phase is characterized by a Luttinger volume that counts
only the $c$ fermions. As the temperature drops the entropy associated with the local moment has to be quenched, 
and competing quenching mechanisms are at the origin of the different phase diagrams. 
Heavy fermion character~\cite{Lohneysen_rev} is associated with the screening of the local moments 
by the conduction electrons. In the framework of the periodic Anderson model, the $f$ fermions 
delocalize so as to a generate the superexchange scale and reappear in the Luttinger sum rule. For our 
odd-sized cluster topologies only this scenario can occur. It is important to 
note, however, that the step from $N_{\text{u}} =1 $ to $N_{\text{u}}=3$ is linked to a substantial decrease of the coherence 
temperature since for the $N_{\text{u}} = 3$ cluster the entropy is partially lifted due to intralayer spin correlations. 

On clusters of even size, the $f$ fermions can form an insulating spin-singlet state and hence quench the entropy without 
involving the first layer fermions. This allows for a band-selective Mott transition---or Kondo breakdown---in 
which the $f$ fermions drop out of the Luttinger volume down to the lowest temperature. Despite
the breaking of translation invariance inherent to the CDMFT, the Luttinger sum rule still holds when 
formulated in the Brillouin zone of the supercell Bravais lattice. 

Given this odd-even effect, the extrapolation to large cluster sizes is difficult and bound to be speculative. One can 
conjecture that for even site lattices, $T_c$ is set by the spin gap $\Delta_{\text{sp}}$ of the corresponding 
half-filled Hubbard model of the first layer. At $U_f/t = 12 $, $\Delta_{\text{sp}}/t = 0.325 $ for $N_{\text{u}}= 2$ whereas 
$\Delta_{\text{sp}}/t =0.214$ for $ N_{\text{u}} = 4$. The decrease in $T_c$ between the $N_{\text{u}}=2$ and $N_{\text{u}} = 4$ clusters is consistent 
with the decrease in the spin-gap. For odd lattices, one can follow the idea that the coherence temperature tracks 
the residual entropy per site of the half-filled Hubbard model on the first layer. Given the above 
conjecture and the fact that the Hubbard model on a triangular lattice 
has a unique ground state, we arrive at the conclusion that the coherence temperature indeed vanishes beyond a 
critical chemical potential. This stands in agreement with the slave boson calculations of 
Ref.~\onlinecite{Benlagra08}. 
If the magnetic system on the first layer orders and breaks a lattice symmetry, then the $f$-fermions can drop out of the Luttinger 
volume without violating the Luttinger theorem. The stability of such a phase with respect to a finite hybridization matrix element generating a Kondo coupling between the layers has been discussed in Ref.~\onlinecite{Yamamoto07}.
On the other hand, if no symmetries are broken such that a spin liquid state 
is realized on the first layer,~\cite{Meng10} 
fractionalized Fermi liquids as proposed in Refs.~\onlinecite{Senthil03,Senthil04} could be realized. 

It is interesting to recast our results in terms of the $Q$-$K$ phase diagram for heavy fermions
proposed by Coleman and Nevidomskyy.~\cite{Coleman10} Here, $K$ 
corresponds to the magnitude of the Kondo screening and $Q$ is a measure of 
frustration between the $f$ fermions. For our model in the local moment regime, the Kondo coupling between the 
two layers is dynamically generated starting from second-order perturbation theory in the hybridization. 
The frustration between the $f$ fermions is generated by the hopping 
matrix element $t_f$, which again in the local moment regime leads to a superexchange interaction between the $f$ fermions. 
In the framework of Ref.~\onlinecite{Coleman10} and in agreement with our numerical simulations, it is the frustration 
between the $f$ fermions that drives the Kondo breakdown or band-selective Mott transition. 
For a recent review in this domain, we refere the reader to Ref.~\onlinecite{Vojta10}. 
This is in contrast to the Kondo lattice model, where the $f$ fermions interact solely through the RKKY interaction, and no 
Kondo breakdown is observed in cluster simulations.~\cite{Martin08,Martin10}

Let us finally return to the bilayer $^3$He experiment. Heavy fermion character is clearly 
seen by the increase of the effective mass (or decrease of the coherence temperature).
Furthermore, and as seen in our calculations, $T_0$ marks the crossover between 
a Curie-Weiss and Pauli behavior of the spin susceptibility. The $Q$ or competing interaction which 
localizes the $f$ fermions to the first layer are the three-body exchange processes. These processes, which 
in solid $^3$He can dominate the two-body antiferromagnetic exchange, lead to 
the observed ferromagnetic behavior. Given this interpretation of the experiment, an extremely important issue 
would be to pin down the experimental value of the Weiss constant. Above $T_0$ it should be positive and essentially 
track the Kondo scale. The transition to the ferromagnetic state should be accompanied by a vanishing and subsequently 
negative value of the Weiss constant.

\acknowledgments

We would like to thank J. Saunders and A. Benlagra for valuable discussions and M. Bercx for a careful reading of the 
manuscript.
The numerical calculations were carried out at the LRZ-M\"unich and
the J\"ulich Supercomputing center. We thank those institutions for their generous allocation of CPU time. 
KSDB thanks the Humboldt foundation for financial support as well as the FFA and DFG 
under grant number AS120/6-1 (FOR1162).


\begin{thebibliography}{10}

\bibitem{Casey03}
A.\ Casey, H.\ Patel, J.\ Ny\'eki, B.\ P.\ Cowan, and J.\ Saunders, 
Phys.\ Rev.\ Lett.\ {\bf 90}, 115301 (2003).

\bibitem{Vollhardt_rev}
D.\ Vollhardt, Rev.\ Mod.\ Phys.\ {\bf 56}, 99 (1984).

\bibitem{Imada_rev}
M.\ Imada, A.\ Fujimori, and Y.\ Tokura, Rev.\ Mod.\ Phys.\ {\bf 70}, 1039 (1998).

\bibitem{Thouless65}
D.\ Thouless, Proc.\ Phys.\ Soc.\ {\bf 86}, 893 (1965).

\bibitem{Roger83}
M.\ Roger, J.\ H.\ Hetherington, and J.\ M.\ Delrieu, Rev.\ Mod.\ Phys.\ {\bf 55}, 1 (1983).

\bibitem{Elser89}
V.\ Elser, Phys.\ Rev.\ Lett.\ {\bf 62}, 2405 (1989).

\bibitem{Neumann07}
M.\ Neumann, J.\ Ny\'eki, B.\ Cowan, and J.\ Saunders, 
Science {\bf 317}, 1356 (2007).

\bibitem{Benlagra08}
A.\ Benlagra and C.\ P\'{e}pin, Phys.\ Rev.\ Lett.\ {\bf 100}, 176401 (2008).

\bibitem{Benlagra09}
A.\ Benlagra and C.\ P\'{e}pin, Phys.\ Rev.\ B {\bf 79}, 045112 (2009).

\bibitem{Biroli04}
G.\ Biroli, O.\ Parcollet, and G. Kotliar, Phys.\ Rev.\ B {\bf 69}, 205108 (2004).

\bibitem{Beach09}
K.\ S.\ D.\ Beach and F.\ F.\ Assaad, arXiv:0905.1127 (2009).

\bibitem{Sandvik98}
A.\ W.\ Sandvik, Phys.\ Rev.\ B {\bf 57}, 10287 (1998).

\bibitem{Beach04a}
K.\ S.\ D.\ Beach, arXiv:0403055 (2004).

\bibitem{Kyung07}
B.\ Kyung, Phys.\ Rev.\ B {\bf 75}, 033102 (2007).

\bibitem{Lohneysen_rev}
H.\ v.\ L\"ohneysen, A.\ Rosch, M.\ Vojta, and P.\ W\"olfle, 
Rev.\ Mod.\ Phys.\ {\bf 79}, 1015 (2007).

\bibitem{Yamamoto07}
S.\ J.\ Yamamoto and Q.\ Si, Phys.\ Rev.\ Lett.\ {\bf 99}, 016401 (2007).

\bibitem{Meng10}
Z.\ Y.\ Meng, T.\ C.\ Lang, S.\ Wessel, F.\ F\. Assaad, and A.\ Muramatsu, 
Nature {\bf 464}, 847 (2010).

\bibitem{Senthil03}
T.\ Senthil, S.\ Sachdev, and M.\ Vojta, 
Phys.\ Rev.\ Lett.\ {\bf 90}, 216403 (2003).

\bibitem{Senthil04}
T.\ Senthil, M.\ Vojta, and S.\ Sachdev, Phys.\ Rev.\ B {\bf 69}, 035111 (2004).

\bibitem{Coleman10}
P.\ Coleman and A.\ H.\ Nevidomskyy, J.\ Low Temp.\ Phys.\ {\bf 161}, 182 (2010).

\bibitem{Vojta10}
M.\ Vojta, J.\ Low Temp.\ Phys.\ {\bf 161}, 203 (2010).

\bibitem{Martin08}
L.\ C.\ Martin and F.\ F.\ Assaad, Phys.\ Rev.\ Lett.\ {\bf 101}, 066404 (2008).

\bibitem{Martin10}
L.\ C.\ Martin, M.\ Bercx, and F.\ F.\ Assaad, arXiv:1007.0010v1 (2010).

\end{thebibliography}
\end{document}